\documentclass[sn-mathphys-num]{sn-jnl}

\usepackage{amsmath,amssymb,amsfonts,graphicx,array}
\usepackage{dcolumn,soul}%

\usepackage{amsthm}
\usepackage{mathrsfs}
\usepackage[figuresright]{rotating}%
\usepackage{algorithm, algorithmicx, algpseudocode}
\usepackage{listings}%
\newtheorem{remark}{Remark}

\begin{document}

\title{A Novel Paradigm in Solving Multiscale Problems}

\author[1]{\fnm{Jing} \sur{Wang}}

\author[1]{\fnm{Zheng} \sur{Li}}

\author[1]{\fnm{Pengyu} \sur{Lai}}

\author[1]{\fnm{Rui} \sur{Wang}}

\author[1]{\fnm{Di} \sur{Yang}}

\author[1]{\fnm{Dewu} \sur{Yang}}

\author*[1,2]{\fnm{Hui} \sur{Xu}}\email{dr.hxu@sjtu.edu.cn}

\author*[3]{\fnm{Wen-Quan} \sur{Tao}}\email{wqtao@mail.xjtu.edu.cn}

\affil*[1]{\orgdiv{School of Aeronautics and Astronautics}, \orgname{Shanghai Jiao Tong University}, \orgaddress{\city{Shanghai}, \country{China}}}
\affil*[2]{Department of Aeronautics, Imperial College London, London, UK}
\affil*[3]{Key Laboratory of Thermo-Fluid Science and Engineering of MOE, School of Energy and Power Engineering, Xi'an Jiao Tong University, Xi'an, China}

\abstract{Multiscale phenomena manifest across various scientific domains, presenting a ubiquitous challenge in accurately and effectively simulating multiscale dynamics in complex systems. In this paper, a novel decoupling solving paradigm is proposed through modelling large-scale dynamics independently and treating small-scale dynamics as a slaved system. A Spectral Physics-informed Neural Network (PINN) is developed to characterize the small-scale system in an efficient and accurate way, addressing the challenges posed by the representation of multiscale dynamics in neural networks. The effectiveness of the method is demonstrated through extensive numerical experiments, including one-dimensional Kuramot-Sivashinsky equation, two- and three-dimensional Navier-Stokes equations, showcasing its versatility in addressing problems of fluid dynamics. Furthermore, we also delve into the application of the proposed approach to more complex problems, including non-uniform meshes, complex geometries, large-scale data with noise, and high-dimensional small-scale dynamics. The discussions about these scenarios contribute to a comprehensive understanding of the method's capabilities and limitations. By enabling the acquisition of large-scale data with minimal computational demands, coupled with the efficient and accurate characterization of small-scale dynamics via Spectral PINN, our approach offers a valuable and promising approach for researchers seeking to tackle multiscale phenomena effectively.}


\maketitle

%% main text
\section{Main} \label{sec:sec1}

% 1. 必要性/重要性+难点
Multiscale phenomena manifest across a wide array of scientific disciplines, spanning from physics~\cite{mahadevan2016impact} and biology~\cite{robinson2005multiscale} to engineering~\cite{weinan2003multiscale} and social systems~\cite{bellomo2011modeling}. 
These phenomena entail intricate patterns and interactions across various spatiotemporal scales, rendering the prediction and analysis of complex systems a formidable challenge~\cite{alber2019integrating, steinhauser2017computational}. 
At the heart of this complexity lies the interplay between large-scale dynamics, which encapsulate dominant patterns and structures, and small-scale dynamics, characterized by finer details and rapid fluctuations~\cite{holmes1996}. 
While large-scale dynamics offer a macroscopic view of system behavior, small-scale dynamics significantly influence the overall system dynamics. 
Accurately predicting and comprehending the full range of spatiotemporal scales not only enhances our capacity to simulate and model diverse phenomena but also unlocks unprecedented insights into emergent properties, facilitating the design and optimization of novel materials, biological processes, and technological systems~\cite{kan2019recent}.

% 2. 已有方法及其优缺点
%% 传统仿真
In the pursuit of unravelling multiscale dynamics in complex systems, traditional numerical methodologies such as in fluid mechanics, Direct Numerical Simulation (DNS)~\cite{moin1998direct}, Large Eddy Simulation (LES)~\cite{mason1994large} and Lattice Boltzmann (LBM) have been widely employed. 
DNS aims to provide detailed insights into the smallest features by providing a comprehensive resolution of all scales in the flow field. Despite significant advancements in high-performance computing, the computational demands associated with DNS increases with the complexity of the system. Consequently, the application of DNS at high Reynolds numbers becomes economically and computationally impractical, especially in scenarios representing both natural and engineering fluid flows~\cite{cant2002high}. 
On the other hand, LES strikes a pragmatic balance by employing sub-grid scale models, resulting in a more computationally feasible approach. Nevertheless, the efficiency of LES also faces challenges, as accurately modelling subgrid-scale interactions remains a persistent issue, potentially introducing inaccuracies in predicting small-scale dynamics~\cite{piomelli1999large}.
Alternatively, the LBM has emerged as a simpler and more parallelizable approach based on the lattice Boltzmann equation on a mesoscopic scale. While LBM excels in processing complex boundaries, its requirement for a large number of lattice numbers with a very small time step makes field acquisition time-consuming and computationally expensive. 
To enhance the small-scale simulations, several important techniques have been developed~\cite{vanden2007heterogeneous,weinan2003heterognous}, such as multigrid methods, domain decomposition methods, adaptive mesh refinement techniques, and multi-resolution methods using wavelets.
These approaches, while focusing computational resources on local regions, may encounter challenges in achieving a substantial enhancement in resolution and computational efficiency across the entire simulation domain~\cite{fish2021mesoscopic}.

%% AI-based
With the development of artificial intelligence~\cite{lecun2015deep}, neural networks (NNs) have shown great potential in the field of fluid dynamics for their abilities to learn complex relationships from data and predict the behaviour of fluid flows~\cite{brunton2019data,wang2023scientific}. One area where NNs have been particularly successful is in the study of 
modelling the Reynolds stress tensor in Reynolds-averaged Navier-Stokes (RANS) simulations~\cite{ling2016reynolds,wangjx2017} and subgrid-scale modelling in LES~\cite{lapeyre2019training}. By leveraging the flexibility and adaptability of NNs, these approaches aim to capture intricate flow behaviours that may be challenging for conventional models to represent accurately.
\cite{buaria2023forecasting} captures the small-scale dynamics of turbulence via the velocity gradient tensor through deep neural networks on existing DNS data at low and moderate Reynolds and demonstrate the capability for predicting their dynamics at both seen and higher unseen Reynolds.
Additionally, the advancement of single-image super-resolution methods~\cite{yang2014single} has partly addressed the challenge of achieving high-resolution reconstruction of flow fields~\cite{vlachas2022multiscale, fukami2019super}. 
However, it is essential to acknowledge the inherent challenges associated with this approach, including the susceptibility to noise amplification and the computational demands involved in processing high-resolution data. 
For practical scientific or engineering problems with definite governing equations, introducing appropriate physics into deep learning algorithms can lead to better solutions for these problems~\cite{peng2021multiscale}.

NNs have also emerged as a novel tool for directly solving partial differential equations (PDEs), to obtain comprehensive flow field solutions. 
Methods like Physics-Informed Neural Networks (PINNs)~\cite{RAISSI2019686}, DGM~\cite{sirignano2018dgm}, Deep Ritz Method (DRM)~\cite{yu2018deep} and PhyCRNet~\cite{rao2023encoding} utilize neural networks to approximate solutions by incorporating governing equations and boundary conditions to the loss function or the neural network architecture. This kind of approach opens avenues for more efficient and accurate solutions to complex fluid dynamics problems and has demonstrated remarkable success in providing efficient solutions for problems ranging from turbulence modelling to the simulation of complex flows~\cite{raissi2020,xu2021explore}. 
Nevertheless, these methods often fail to solve challenging PDEs when the solution exhibits high-frequency or multiscale structure due to the `spectral bias', as highlighted in prior works~\cite{cao2020understanding,tancik2020fourier,basri2020}.
While some efforts have been made to enhance multiscale simulation by imposing multiscale priors into the network~\cite{wang2021eigenvector,kim2024review}, improvements remain limited. 
In addition to solution approximation, another category of neural network applications involves solution mapping between infinite-dimensional function spaces. Approaches such as DeepOnet~\cite{lu2021learning}, Fourier Neural Operator (FNO)~\cite{li2020fourier} and PDE-Net~\cite{long2018pde} are proposed to learn to solve an entire family of PDEs. These methods can approximate complex operators raising in PDEs that are highly non-linear and have demonstrated promising results across a wide range of applications~\cite{wang2021flow}.
While these methods hold promise for achieving or accelerating multiscale simulations~\cite{bhatia2021machine,tao2010nonintrusive,kevrekidis2003equation}, challenges still arise in terms of limited generalization and computational intensity.

% 3. 本文目的及方法
Inspired by the scale decomposition principles from adiabatic elimination\cite{van1985elimination}, classical center-manifold theory\cite{mercer1990centre, linot2020deep}, Mori-Zwanzig formalism\cite{sture1974strategies} and the extended evolutionary renormalization group (RG) method\cite{schmuch2013}, 
we develop a novel decoupling solving mode to simulate the multiscale dynamics through modelling the large-scale dynamics independently and treating the small-scale dynamics as a slaved system.
The large-scale dynamics can be resolved through modern low computational demands, such as coarse meshes or Reynolds-Averaged Navier-Stokes (RANS) simulations.
The PINN is selected as the fundamental architecture to characterize the small-scale dynamics based on the large-scale dynamics for their capability to address limitations associated with conventional methods.
Unlike traditional approaches, PINNs operates without the reliance on predefined meshes, showcasing natural adaptability to varying scales within the system. This intrinsic mesh independence is particularly valuable when dealing with complex fluid dynamics where small-scale features demand precise representation.
Furthermore, PINNs offer a significant advantage in terms of computational efficiency. By mitigating the need for elaborate meshes, these neural networks can substantially reduce the associated computational expenses, making them a cost-effective alternative to traditional methods. This efficiency, however, does not compromise accuracy, as PINNs maintain a high level of precision in capturing the intricate details of the underlying physics.
To mitigate the challenges related to the `spectral bias' in PINNs, the equation for solving the small-scale dynamics is derived and directly learned in the spectral space. In our investigation, the Fourier space is taken for representing functions at different scales~\cite{wang2021eigenvector}, which is leveraged to overcome the limitations and improve the overall performance of the model.
Throughout this article, we demonstrate the capabilities of the proposed method by applying it to extensive numerical experiments.
This decoupling approach simplifies the analysis and prediction of spatiotemporal systems, enabling a deep understanding of interactions between different scales and facilitating more efficient computational simulations.

\section{Results}
% PDE
We consider the general classical form of PDEs in an operator form with a finite time horizon $T\in \mathbb{R}^+$ and a bounded domain $D\subset \mathbb{R}^d$:
\begin{equation}\label{eq:op}
\left\{\begin{aligned}
\displaystyle \frac{\partial u}{\partial t}&=\mathscr{A}(u),  && \text{in}\ [0,T]\times D, \\
u&=g,&& \text{in}\  [0,T]\times \partial D, \\
u&=u_0,&& \text{in}\ \{0\}\times D,
\end{aligned}
\right.
\end{equation}
where $u: [0,T]\times D\rightarrow \mathbb{R}^{\bar d}$ is the solution and $u(t,\cdot)$  belongs to a function space $\mathscr{X}$. $\mathscr{A}$ is a nonlinear partial differential operator with domain $\mathscr{D}(\mathscr{X})\subset \mathscr{X}$. $u_0$ and $g$ denote the initial and boundary conditions, respectively.  

% definition of large/small-scale
%Generally, the solution $u$ is approximated in a finite-dimensional space by numerical computation.  
The function space $\mathscr{X}$ is an infinite-dimensional space and Eq.~\eqref{eq:op} usually can't be solved accurately in this space. Consider $\mathscr{X}^N$ as a finite-dimensional subspace of $\mathscr{X}$, and the function space $\mathscr{X}$ has the following decomposition $\mathscr{X}=\mathscr{X}^N\oplus(\mathscr{X}/\mathscr{X}^N).$ If the solution $u(t,\cdot)$ is well-approximated in $\mathscr{X}^N$, the space $\mathscr{X}/\mathscr{X}^N$ can be negligible. Therefore, the solution discussed in this paper is approximated in $\mathscr{X}^N$ and it is consistently denoted as $u(t,x)$ for the sake of simplicity throughout the manuscript.

While $\mathscr{X}^N$ is already a finite-dimensional space, it's crucial to choose $N$ large enough to obtain a sufficiently accurate solution, which may still incur high computational costs. Let $\mathscr{X}^M$ be a finite-dimensional subspace of $\mathscr{X}^N$ that captures the dominant behaviour or large-scale patterns of the solution. Simultaneously, introduce a complementary space ${\mathscr{X}^M}^\perp$ to account for small-scale behaviours and serve as an additional correction. Furthermore, the finite-dimensional space can be decomposed as:  
\begin{equation}
\mathscr{X}^N=\mathscr{X}^M\oplus {\mathscr{X}^M}^\perp, 
\end{equation}
where $M<N$. Then we could use different methods to solve the problem on $\mathscr{X}^M$ and ${\mathscr{X}^M}^\perp$  respectively and obtain a high-fidelity solution with lower computational costs.

% equation
Let $\mathscr{P}_M$ and $\mathscr{Q}_M$ denote the projection operator from $\mathscr{X}$ to ${\mathscr{X}^M}$ and ${\mathscr{X}^M}^\perp$, respectively.  Then the well-resolved solution can be expressed as $u=p_M+q_M$ with $p_M=\mathscr{P}_M u$ and $q_M=\mathscr{Q}_M u$. Essentially, $p_M$ and $q_M$ are synergistic. If we project Eq.~\eqref{eq:op}, respectively, onto the large-scale space and small-scale space, we obtain: 
\begin{equation}\label{eq:ls}
\frac{\partial p_M}{\partial t}=\mathscr{P}_M\mathscr{A}(p_M+q_M),
\end{equation}
\begin{equation}\label{eq:ss}
\frac{\partial q_M}{\partial t}=\mathscr{Q}_M\mathscr{A}(p_M+q_M).
\end{equation}
According to Eq.~\eqref{eq:ss}, $q_M$ can be treated as parasitic structures from $p_M$, which helps to capture the fine-scale details and accurately represent the solution behaviour. The above belongs to the classical realm of the nonlinear Galerkin method used to describe the decomposition of a complex system in the sense of slow-and-fast modes~\cite{foias1989exponential}. Based on this solution decomposition, it is possible to obtain the dominant dynamics and the correction term of the solution in a decoupled way, which can simplify the problem, enable a deep understanding of interactions between different scales, and make it more computationally feasible. Generally, the dynamics of $p_M$ can be easily and properly captured even when the term related to $q_M$ is neglected in Eq.~\eqref{eq:ls}. 
Therefore, this paper develops a spectral PINN method to obtain $q_M$ by Eq.~\eqref{eq:ss} with given $p_M$, which establishes a general computationally feasible way to characterize the small-scale dynamics accurately and efficiently.

\begin{figure*}
\centering
\includegraphics[width=1.\textwidth]{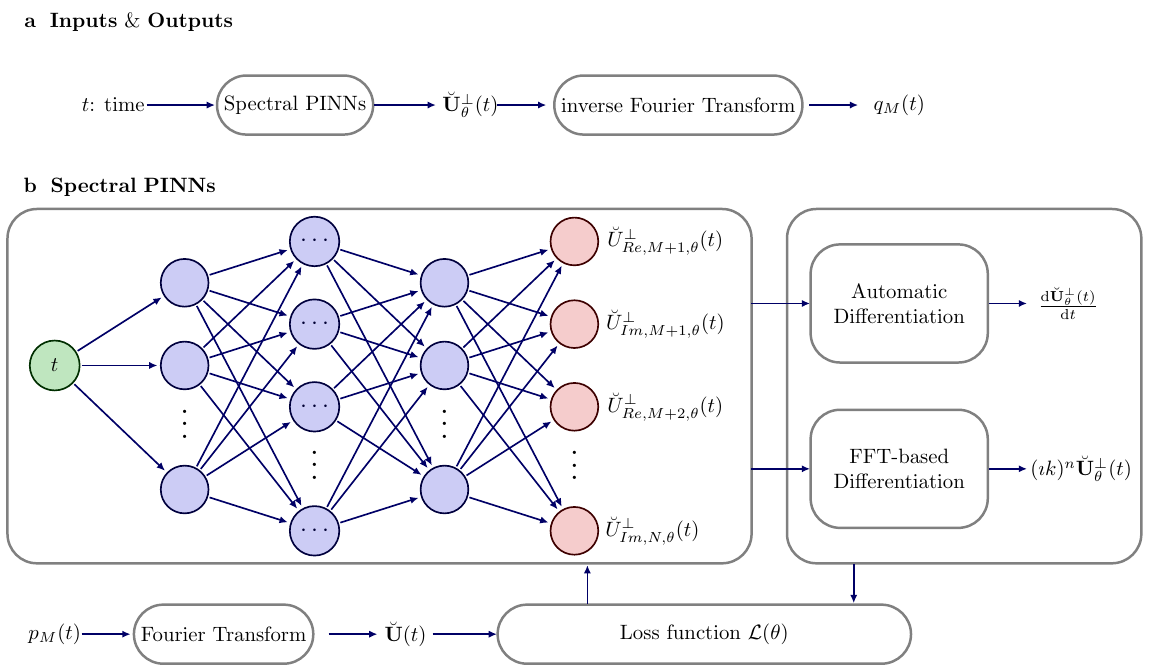}
\caption{\textbf{llustrations of the problem set-up and schematic diagram of the spectral PINN configuration}. 
\textbf{a}. The network takes time as input and predicts the Fourier coefficient of the small-scale data, followed by an inverse Fourier Transform.
\textbf{b}. Schematic diagram of the spectral PINN method.}
\end{figure*}

\subsection{Case 1: One-dimensional Kuramoto–Sivashinsky equation}

% The definition of ks equation
The Kuramoto–Sivashinsky (KS) equation holds a pivotal position in the realm of nonlinear partial differential equations, particularly within the study of spatiotemporal chaos and pattern formation. Its unique ability to capture complex patterns and intricate spatiotemporal dynamics makes it a valuable tool for investigating flame front stability, reaction-diffusion systems, and various other physical contexts. Characterized by fourth-order spatial derivative and nonlinear terms, we consider the following KS equation in one space dimension with periodic boundary conditions:
\begin{equation}\small\label{eq:ks}
\begin{aligned}
\frac{\partial u}{\partial t} + u\frac{\partial u}{\partial x} + \frac{\partial^2 u}{\partial x^2} + \frac{\partial^4 u}{\partial x^4}=0,  &&\text{in}\ [0,T]\times [L_1,L_2], \\
u(x, 0)=u_0(x), &&\text{in}\{0\}\times [L_1,L_2] ,
\end{aligned}
\end{equation}
where $u\in C([0,T];H_{\mathrm{per}}^r((L_1,L_2);\mathbb{R}))$ with $r>0$, $u_0\in L((L_1,L_2);\mathbb{R})$ is the initial condition.
% The numerical solution of ks1
To generate the numerical solution as the ground-truth reference, the equation is solved by the pseudo-spectral method combined with the exponential time-differencing fourth-order Runge–Kutta formula. In particular, we start from a simple flow with the initial condition $u(x,0) = -\sin(\pi x/10),x\in[-10,10]$. The time integration is implemented until the final time $T = 50$s with a time-step size of 0.1.  The spatial discretisation is implemented with 512 Fourier modes. 
To generate large-scale data specially,  the Fourier transform is applied to the entire dataset, yielding large-scale data characterized by 4 modes, and small-scale dynamics represented by 10 modes. Prediction models are then constructed based on the large-scale data. Significantly, the energy distribution is quantified from the numerical results, revealing that 89.98\% of the total energy is attributed to the large-scale components, while the remaining 10.02\% is associated with the small-scale components.
We let the small-scale solution be represented by a 6-layer deep neural network, each followed by a GELU activation function. The input layer takes one feature, and subsequent hidden layers have 100, 300, 500, and 300 neurons, respectively, before reaching the output layer with 20 neurons which represent the real and imaginary parts of the 10 modes, respectively. The network is trained for 5,000 epochs using the Adam optimizer~\cite{kingma2014adam}. The mean squared error of the residuals of the equation and initial condition is used as the optimization objective for obtaining a set of model parameters. The Cosine learning rate warmup~\cite{gotmare2018closer} is employed, where the learning rate increases linearly from 0 to 0.008 over 50 epochs and decreases from 0.008 to 0 over the remaining epochs following a cosine curve.

\begin{figure*}
\centering
\includegraphics[width=1.\textwidth]{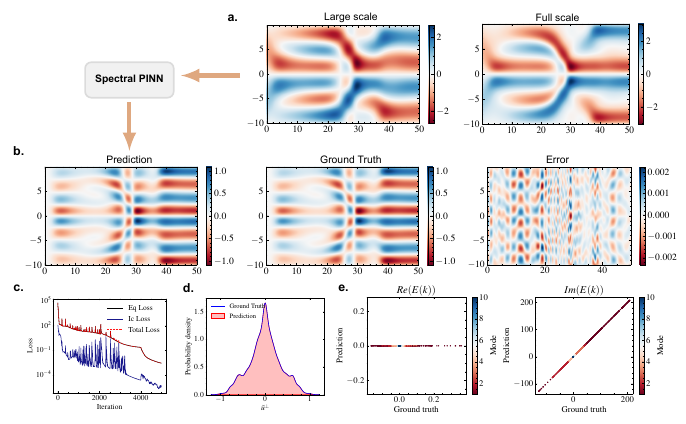}
\caption{\textbf{KS equation with the initial condition $u(x,0) = -sin(\pi x/10)$}: 
\textbf{a}. Comprehensive depiction of the full-scale field and extracted large-scale field. 
\textbf{b}. Comparative analysis of the small-scale dynamics derived from the numerical simulation and the proposed method. 
\textbf{c}. The training trajectory of the network.
\textbf{d}. Comparison of the PDFs of the obtained small-scale fields.
\textbf{e}. Comparative assessment of the real and imaginary components of Fourier coefficients signifying the small-scale dynamics, with the mode-specific colouration of data points.}
\label{fig:ks1}
\end{figure*}

% the results for solving ks1
The detailed results are presented in Fig.~\ref{fig:ks1}. It is seen that the training errors converge with an increase in training steps. After 5000 epochs, the error associated with the PDE equation is minimized to 0.0027, reflecting the model's ability to satisfy the underlying equations. The error of the initial condition is impressively low at 0.0004, indicating the model's proficiency in reproducing the specified initial conditions. 
Specifically, a comparison is made between the Fourier coefficients of the learned small-scale solutions and their exact counterparts. The observed match in the imaginary part with the reference is noteworthy. While a marginal error is noted in the real part for this particular case, the accuracy remains within acceptable bounds.
Consequently, the predicted contour patterns and the probability density functions (PDFs) of the fields demonstrate a satisfactory match with the exact results. The mean average error over the solution is around 0.0003, indicating an excellent fit to the exact solution. 
The surprising result strongly indicates that the algorithm is accurately learning the underlying partial diﬀerential equation of the small-scale dynamics.

% The settings of the chaotic KS2
It is important to note that the above study deliberately avoids regions where the KS equation exhibits chaotic behaviour, of which the small-scale dynamics are relatively simple to obtain. To better reveal the proposed method, an exploration into a chaotic flow is considered by expanding the spatial domain to $[0, 32\pi]$ and altering the initial condition to $u(x,0) = \cos(x/16)(1 + \sin(x/16))$. The equation is numerically integrated over the extended spatial domain up to the final time  $T = 100$s with time-step size 0.1, and implementing 1024 Fourier modes for spatial discretisation. This expanded computational setup allows for the generation of intricate solutions.
In this case, the large-scale data is characterized by 15 modes, and the accurate representation of chaotic small-scale dynamics with 20 modes is undertaken for prediction based on the large-scale data. The energies of the large and small scales are 76.74\% and 23.23\%, respectively. The solution is represented by the deep neural network that shares the same architecture in a 6-layer deep neural network, each layer followed by a GELU activation function. The input layer processes one feature, and subsequent hidden layers consist of 100, 300, 500, and 300 neurons, respectively, before reaching the output layer with 40 neurons which represent the real and imaginary parts of the 20 modes, respectively. The network undergoes training for 5,000 epochs using the Adam optimizer. The mean squared error of the residuals of the equation and initial condition is used as the optimization objective for obtaining a set of model parameters. The Cosine learning rate warmup is also implemented, with the learning rate linearly increasing from 0 to 0.008 over 50 epochs and then decreasing from 0.008 to 0 over the remaining epochs following a cosine curve.

\begin{figure*}
\centering
\includegraphics[width=1.\textwidth]{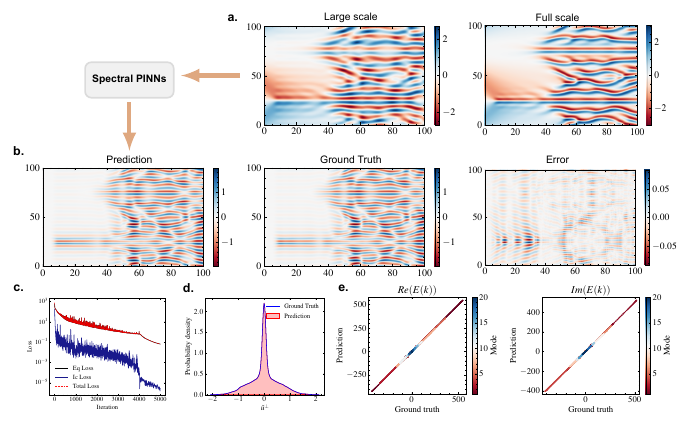}
\caption{\textbf{KS equation with the initial condition $u(x,0) = \cos(x/16)(1 + \sin(x/16))$}: 
\textbf{a}. Comprehensive depiction of the full-scale field and extracted large-scale field. 
\textbf{b}. Comparative analysis of the small-scale dynamics derived from the numerical simulation and the proposed method. 
\textbf{c}. The training trajectory of the network.
\textbf{d}. Comparison of the PDFs of the obtained small-scale fields.
\textbf{e}. Comparative assessment of the real and imaginary components of Fourier coefficients signifying the small-scale dynamics, with mode-specific colouration of data points.}
\label{fig:ks2}
\end{figure*}

% the results for solving ks2
Fig.~\ref{fig:ks2} shows the detailed results. It is evident that, as the number of training steps increases, the training errors converge. After 5,000 epochs, the errors of the PDE equation and initial condition are decreased to 0.064 and 0.0000028, respectively.  
A meticulous comparison is also conducted between the Fourier coefficients of the acquired small-scale solutions and their exact counterparts. 
Specifically, the Fourier coefficients and whole flow field of the learned and exact chaotic small-scale solutions are compared and visualized.  The error of the Fourier coefficients converges at 1.16, indicating the model's effective capture of the chaotic Fourier modes in the small-scale data. It is observed that, whether the predicted contour patterns or the PDFs, exhibit a commendable congruence with the exact results and all the fine structures are all identically reproduced. 
In the meanwhile, the mean average error of the prediction, which is around 0.008, is much smaller compared with the scale of the ground truth, showing the accuracy of the proposed method as a PDE solver.

\begin{figure*}
\centering
\includegraphics[width=1.\textwidth]{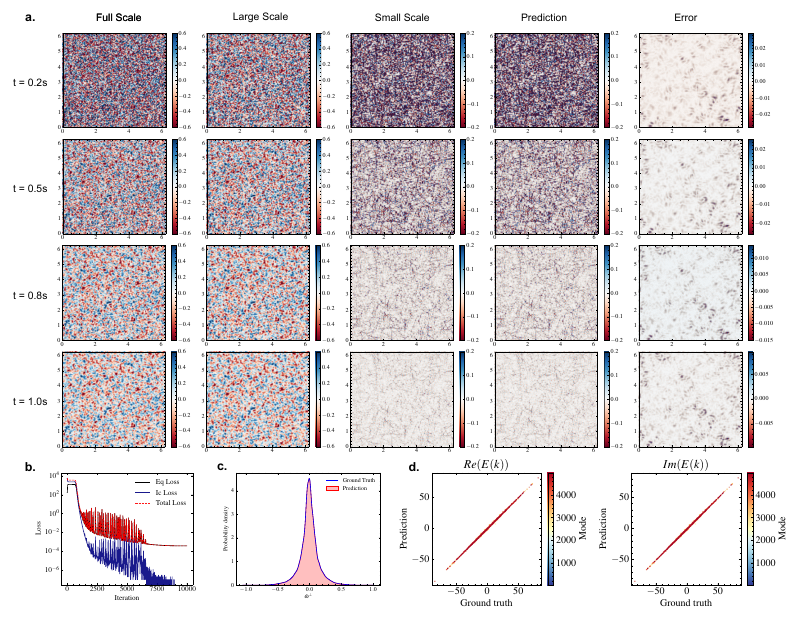}
\caption{\textbf{2D NS equations}:
\textbf{a}. A comprehensive depiction of the full-scale, extracted large-scale and small-scale fields derived from the numerical simulation, along with the predicted small-scale dynamics obtained with the proposed method. The detailed results at each time step are provided in the Supplementary Video 1.
\textbf{b}. The training trajectory of the network.
\textbf{c}. Comparison of the probability density of the obtained small-scale fields.
\textbf{d}. Comparative assessment of the real and imaginary components of Fourier coefficients signifying the small-scale dynamics, with mode-specific colouration of data points.}
\label{fig:ns2d}
\end{figure*}

\subsection{Case 2: Two-dimensional Navier-Stokes equations}
% The definition of 2d ns equation
As the basic equation for modelling the weather, ocean currents, water flow in a pipe and airflow around a wing, we consider the following two-dimensional (2D) viscous, incompressible Navier-Stokes (NS) equations:
\begin{equation}
\begin{aligned}
\partial_t \omega+\boldsymbol{u} \cdot \nabla \omega & =\nu \nabla^2 \omega, && \text{in} [0,T]\times [0,L]^2 \\
\omega&=\omega_0,  && \text{in}\ \{0\}\times [0,L]^2 \\
\end{aligned}
\end{equation}
where $\boldsymbol{u} \in C\left([0, T] ; H_{\mathrm{per}}^r\left((0,L)^2 ; \mathbb{R}^2\right)\right)$ for any $r>0$ is the velocity field, $\omega=\nabla \times \boldsymbol{u}$ is the vorticity, $w_0 \in L_{\mathrm{per}}^2\left((0,L)^2 ; \mathbb{R}\right)$ is the initial vorticity, $\nu \in \mathbb{R}_{+}$is the viscosity coefficient.
We consider the system with $\nu=5.0\times 10^{-4}$. 
The time integration is implemented until the final time $T = 1$s with time-step size 0.001. 
With $L = 2\pi$, 128 Fourier modes are employed to implement the spatial discretisation along $x$ and $y$ directions.
The initial condition of the problem is randomly sampled while periodic BC is adopted.
The equation is solved by the pseudo-spectral method combined with the Crank-Nicholson method. 
In this equation, large-scale data is systematically acquired with a cut-off radius set at 60. Remarkably, an in-depth analysis of the energy distribution, based on numerical results, reveals a substantial concentration of 83.97\% of the total energy within the large-scale components. The small-scale data targeted for prediction corresponds to those with cut-off radii exceeding 60 but falling below 85, with 15.78\% of the energy concentrated in the resulting small-scale components.  The neural network, representing the solution $u_s$, is designed as a 6-layer deep architecture with each layer followed by a GELU activation function. The input layer accommodates a single feature, followed by hidden layers with 100, 500, 1000, 2000, and 9956 neurons, respectively. The network undergoes training for 10,000 epochs using the Adam optimizer. The chosen loss function is MSELoss. The learning rate undergoes a Cosine learning rate warm-up, initiating with a linear increase from 0 to 0.0006 over 100 epochs and concluding with a decrease from 0.0006 to 0 over the remaining epochs, following a cosine curve.

% Results 
The convergence results obtained with the proposed method are depicted in Fig.~\ref{fig:ns2d}. 
After training, the MSE of the model drops to 0.56. 
To illustrate the eﬀectiveness of our approach, Visualization of the learned small-scale dynamics in conjunction with the exact solution is also presented. 
The mean average error of the overall solution is about 0.002, which is much smaller compared with the scale of the ground truth. 
Absolute differences in the figure show that our model performs well in solving the small-scale equation. The model performs sufficiently well in capturing overall flow features. 
Particularly noteworthy is the comparison of Fourier coefficients and PDFs between the learned and exact small-scale solutions showcased, which presents an outstanding fit for the exact solution.

\subsection{Case 3: Three-dimensional Navier-Stokes equations}
Due to the curse of dimensionality, solving PDEs on three dimensional (3D) spatial domain is considered to be more challenging than the 2D problems. In this section, we explore the scalability of the proposed method on the following 3D Navier-Stokes equations:
\begin{equation}\small
\begin{aligned}
\partial_t \boldsymbol{u}+\boldsymbol{u} \cdot \nabla \boldsymbol{u} & = -\nabla p + \dfrac{1}{Re} \nabla^2 \boldsymbol{u}, & & \text{in}\ [0,T]\times [0,L]^3 \\
\nabla \cdot \boldsymbol{u} &= 0, & & \text{in}\ [0,T]\times [0,L]^3 \\
\boldsymbol{u}&=\boldsymbol{u}_0,  & & \text{in}\ \{0\}\times [0,L]^3 &\\
\end{aligned}
\end{equation}
where $\boldsymbol{u} \in C\left([0, T] ; H_{\mathrm{per}}^r\left((0,L)^3 ; \mathbb{R}^3\right)\right)$ for any $r>0$ is the velocity field. 
$\boldsymbol{u}_0 \in L_{\mathrm{per}}^2\left((0,L)^3 ; \mathbb{R}^3\right)$ is the initial velocity.
To characterize the small-scale flow, we consider the system with Taylor-Reynolds number $Re_{\lambda}=950$. The initial condition is also randomly generated while periodic BC is adopted.
To generate high-fidelity numerical results for comparison, the equation is solved by the pseudo-spectral method with the following transformed form, where the pressure is eliminated~\cite{iovieno2001new}:
\begin{equation}
\partial_t \boldsymbol{U} = -\left( \boldsymbol{I}-\frac{\boldsymbol{k}\boldsymbol{k}^\mathrm{T}}{\boldsymbol{k}^\mathrm{T}\boldsymbol{k}}\right) \cdot \mathscr{F} (\boldsymbol{u} \cdot \nabla \boldsymbol{u} ) - \dfrac{1}{Re} \boldsymbol{k}^\mathrm{T}\boldsymbol{k}  \boldsymbol{U},
\end{equation}
Where $\boldsymbol{U}=\mathscr{F}(\boldsymbol{u})$ and $\mathscr{F}(\boldsymbol{u} \cdot \nabla \boldsymbol{u})$ is obtained by pseudo-spectral real space evaluation.
With $L = 2\pi$, 64 Fourier modes are employed to implement the spatial discretisation along $x$, $y$ and $z$ directions.
The temporal discretisation employed in this study adheres to a Kolmogorov time scale of less than 0.1s, thereby dictating a time interval of 0.2s for solution convergence, covering the temporal span from 9s to 9.2s with a time-step magnitude of 0.001. 
The Kolmogorov time scale for this case is less than 0.1s, thus guiding our selection of a time interval of 0.2s for analysis, spanning from 9s to 9.2s with a time-step size of 0.001s, to effectively characterize the small-scale dynamics.
Acquisition of large-scale data is systematically executed within a prescribed cut-off radius of 8, revealing from numerical analyses a notable aggregation of 87.61\% of the total energy within the large-scale components. Concurrently, the small-scale data, earmarked for predictive modelling, encompasses instances with a cut-off radius surpassing 8 but remaining below 16, with a discernible 10.87\% energy concentration within the resultant small-scale components. 
The neural network, representing the solution corresponding to the small-scale behaviour, is designed as a 5-layer deep architecture with each layer followed by a GELU activation function. The input layer accommodates a single feature, followed by hidden layers with 5000, 10000 and 20000 neurons, respectively, culminating in the output layer with 89772 neurons. The network undergoes training for 10,000 epochs utilizing the Adam optimizer, with the mean square loss as the loss function. We start with a learning rate of 0.0002 and ReduceLROnPlateau~\cite{al2022scheduling} scheduler is used during the training phase to dynamically update the learning rate based on the total loss.

\begin{figure*}
    \centering
    \includegraphics[width=1.\textwidth]{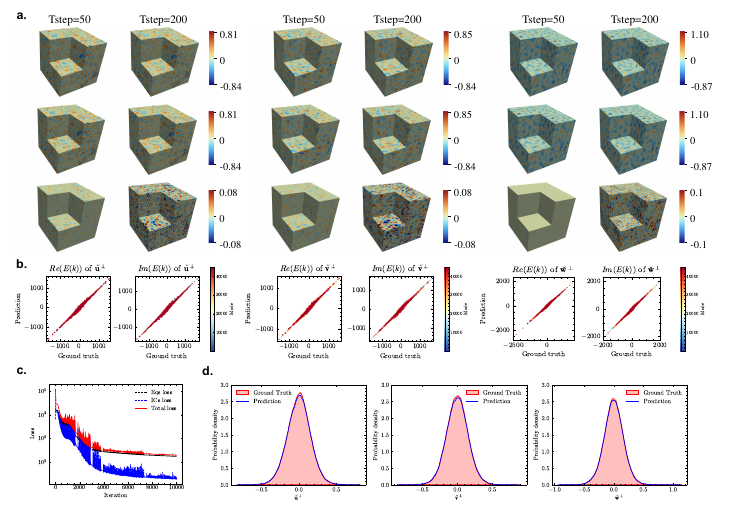}
    \caption{\textbf{3D NS equations}:
\textbf{a}. Comparative analysis of the small-scale dynamics derived from the numerical simulation and the proposed method. The detailed results at each time step are provided in the Supplementary Video 2.
\textbf{b}. Comparative assessment of the real and imaginary components of Fourier coefficients signifying the small-scale dynamics, with mode-specific colouration of data points.
\textbf{c}. The training trajectory of the network.
\textbf{d}. Comparison of the probability density of the obtained small-scale fields.}
    \label{fig:ns3d}
\end{figure*}

% Results 
The convergence and performance of the proposed method for solving the 3D Navier-Stokes equations are comprehensively depicted in Fig.~\ref{fig:ns3d}. 
As the trajectories of loss convergence suggested, the training errors exhibit convergence with the number of training steps increasing.  
To highlight the significance of our approach, we present visual representations of the learnt small-scale dynamics in comparison to the exact solution. The mean average error for the entire solution is less than 0.01, demonstrating the model's precision in predicting complex flow dynamics. The absolute differences depicted in the figures demonstrate the model's ability to resolve small-scale equations while still accurately reflecting overall flow features. 
The model's capacity to predict even minute changes in the flow structure is emphasized. The comparison of the Fourier coefficients and PDFs of the predicted small-scale solutions to their precise counterparts is particularly significant. 
Although the extent of adherence does not reach the level of precision observed in prior instances, the observed deviations predominantly manifest in proximity to the zero velocity regime, exerting minimal influence on the overarching field.
This scrutiny reveals an amazing conformity to the exact solution, increasing trust in the model's correctness and reliability.

\begin{figure*}
    \centering   \includegraphics[width=1.\textwidth]{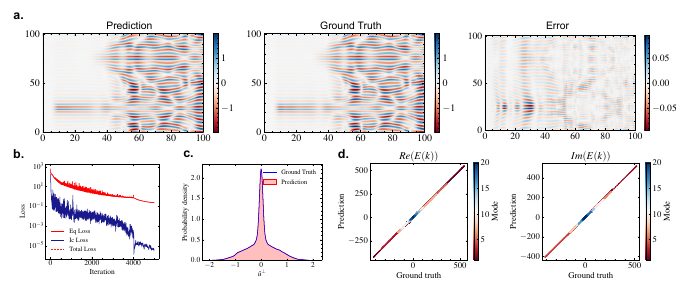}
    \caption{\textbf{KS equation with 10\% noise on the large-scale under the initial condition $u(x,0) = cos(x/16)(1 + sin(x/16))$}: 
\textbf{a}. Comparative analysis of the small-scale dynamics derived from the numerical simulation and the proposed method. 
\textbf{b}. The training trajectory of the network.
\textbf{c}. Comparison of the PDFs of the obtained small-scale fields.
\textbf{d}. Comparative assessment of the real and imaginary components of Fourier coefficients signifying the small-scale dynamics, with the mode-specific colouration of data points.}
\label{fig:ks2noise}
\end{figure*}

\begin{figure*}
    \centering
    \includegraphics[width=1.\textwidth]{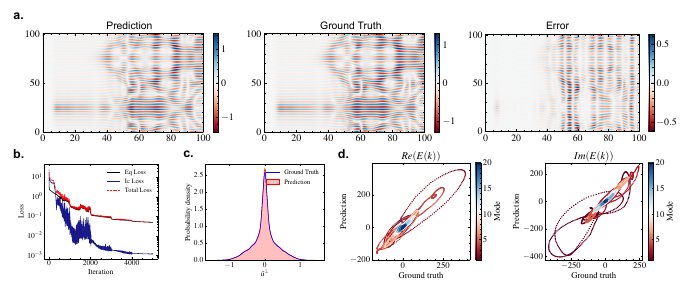}
    \caption{\textbf{KS equation without unsteady terms under the initial condition $u(x,0) = cos(x/16)(1 + sin(x/16))$}: 
\textbf{a}. Comparative analysis of the small-scale dynamics derived from the numerical simulation and the proposed method. 
\textbf{b}. The training trajectory of the network.
\textbf{c}. Comparison of the PDFs of the obtained small-scale fields.
\textbf{d}. Comparative assessment of the real and imaginary components of Fourier coefficients signifying the small-scale dynamics, with the mode-specific colouration of data points.}
\label{fig:kst}
\end{figure*}

\section{Discussion}

In this study, we solely aim to demonstrate the method's potential, without incorporating advanced strategies for PINNs, such as adaptive loss~\cite{xiang2022self} or gradient-enhanced techniques~\cite{yu2022gradient}.
Nevertheless, the effectiveness of our proposed method has been demonstrated in tackling fluid dynamics problems with uniform meshes and regular geometries. The choice of Fourier transform has provided a structured representation that captures the frequency content of the solution. While these cases serve as foundational examples, our method's adaptability extends beyond such straightforward scenarios. In our subsequent discussions, we will delve into the application of our approach to more complex problems, addressing challenges associated with non-uniform meshes, complex geometries, large-scale data with noise, and high-dimensional small-scale dynamics.

To address \textbf{non-uniform meshes}, various strategies can be employed to enhance the adaptability of our proposed method. One effective approach involves utilizing Chebyshev Polynomial expansions, which are particularly well-suited for irregularly spaced points. Chebyshev polynomials offer advantages in accurately representing functions on non-uniform grids, providing a robust solution for spatially varying data. 
Alternatively, PINNs present a modern and versatile solution to transform non-uniform meshes into uniform ones. PDE loss that enforces the governing equations on the large-scale dynamics and data losses that ensure consistency with observed measurements are incorporated. This integration of PINNs allows for adaptive mesh transformations, enabling the model to effectively learn and represent complex spatial dependencies, making our method adaptable to a wide range of mesh configurations in fluid dynamics simulations.

For dealing with \textbf{complex geometries}, two possible solutions are listed here. Firstly, the use of domain decomposition techniques can be employed, where the complex geometry is broken down into simpler subdomains where the proposed method can be applied more easily. This allows for a step-by-step approach to solving complex problems. Secondly, employing advanced mesh generation algorithms, such as adaptive mesh refinement, can help in discretizing complex geometries more effectively. These algorithms dynamically adjust the mesh resolution based on the solution's behaviour, allowing for a more accurate representation of the underlying physical phenomena in intricate geometries. Combining these strategies with the proposed method opens up avenues for tackling a wide range of complex problems efficiently and accurately.

In scenarios with \textbf{large-scale data corrupted by noise}, our proposed method demonstrates a degree of robustness, as evidenced by its application to the KS equation with synthetic measurements of large-scale data accompanied by 10\% Gaussian noise. The learned small-scale dynamics and  Fourier coefficients in conjunction with the exact solution of the 2D NS equation are presented in Fig.~\ref{fig:ks2noise}. 
Despite the presence of noise, the method exhibits resilience, with the predicted dynamics maintaining key structures, even though the overall accuracy may be slightly compromised compared to a noise-free scenario. This resilience suggests that our approach can effectively handle noise-induced uncertainties in large-scale data.
To further address noise-related challenges, alternative solutions can be explored. Denoising techniques, such as signal processing algorithms or machine learning-based approaches specifically designed for noise reduction, can be applied to pre-process the large-scale data. Additionally, incorporating advanced regularization methods in the training process can mitigate the impact of noise on the learning process.

Considering \textbf{the number of modes for small-scale dynamics}, it can be extremely high in complex cases, thus making direct modelling computationally demanding. In our cases, we employ a strategy of truncation by discarding higher-order modes that contribute minimally to the overall energy of the system. This selective neglect of modes with negligible energy allows for a substantial reduction in computational complexity without compromising the accuracy of the simulation. 
To further enhance the representation of complex small-scale features, alternative strategies with improved network architectures can be explored. Hierarchical models allow for a multi-level representation of features, capturing both global and local characteristics. Decomposed networks, on the other hand, can separate the modelling of different modes, providing a modular approach to handle the complexity of high-dimensional small-scale dynamics in fluid systems.

Addressing challenges associated with \textbf{uncertain or hard-to-specify initial conditions} is an integral aspect of our study. Notably, we have identified a viable solution in cases where the relaxation time of the linear part in Eq.~\eqref{eq:ls} is significantly smaller than that of Eq.~\eqref{eq:ls}. In such instances, a reliable approximation emerges as $\mathscr{Q}_M\mathscr{A}(p_M+q_M)=0$. %~\cite{}. 
It is noteworthy that the unsteadiness observed in the small-scale dynamics predominantly stems from the large-scale, rather than being an inherent characteristic. This is justified by the observation that the evolution of the small-scale, induced by itself, is markedly smaller than other contributing terms. The efficacy of this solution is demonstrated through testing on the chaotic KS case, where the learned small-scale dynamics and Fourier coefficients, coupled with the exact solution, are presented in Fig.~\ref{fig:kst}. While overall accuracy may experience a slight compromise compared to the unsteady scenario, particularly evident within the Fourier space, the proposed method demonstrates notable robustness, preserving essential structural features in the predicted dynamics. 
Despite the inclusion of unsteady terms, the proposed method demonstrates notable robustness, preserving essential structural features in the predicted dynamics.

\section{Conclusions}
This paper introduces a novel approach for characterizing small-scale dynamics in complex systems through the development of a decoupling method. By modelling large-scale dynamics independently and treating small-scale dynamics as a slaved system, a Spectral PINN is proposed to approach the small-scale system in an orthogonal basis function space, specifically utilizing Fourier space in this paper. The proposed method introduces a new solving mode, where large-scale data is initially obtained with relatively low computational demands, such as coarse meshes or Reynolds-Averaged Navier-Stokes (RANS) simulations. Subsequently, the proposed Spectral PINN is employed to capture detailed structures of the small-scale dynamics, bringing distinct benefits for improving efficiency and accuracy.
The effectiveness of the method is demonstrated through extensive numerical experiments, ranging from one-dimensional Kuramoto–Sivashinsky equations to three-dimensional Navier-Stokes equations, showcasing its versatility in addressing fluid dynamics problems. Furthermore, the paper delves into the application of the proposed approach to more complex problems, including non-uniform meshes, complex geometries, large-scale data with noise, and high-dimensional small-scale dynamics. 
The results and insights obtained from these scenarios contribute to a comprehensive understanding of the method's capabilities and limitations.
In summary, the development of the method takes advantage of both deep learning and first principles while avoiding the drawbacks associated with exclusively data-driven systems, expanding the possibilities for reliably and effectively predicting complicated multiscale systems.

\section{Methodology}

% whole idea
As a typical PDE, Eq.~\eqref{eq:ss} can certainly be solved through traditional numerical methods. However, extremely fine grids are required, which leads to prohibitively high computational costs. To mitigate this challenge and efficiently address small-scale phenomena, the use of Physics-Informed Neural Networks (PINNs) presents a promising alternative, which leverages the power of deep learning to approximate complicated, nonlinear problems. PINNs offer a cost-effective approach for solving both large-scale and small-scale flow simulations since their computational cost is independent of grid resolution. It makes PINNs a particularly attractive option for tackling small-scale phenomena. In this section, we offer a brief introduction to PINNs and present a detailed overview of our proposed spectral PINN, specifically designed to solve Eq.~\eqref{eq:ss}.
%is proposed to solve the Eq.~\eqref{eq:ss}. The detailed methodology is introduced in this section.
% Establishing slaved dynamics representation by neural network stems from the theory of the nonlinear Galerkin approach.

\subsection{Classical PINN}
% pinns
Before introducing the proposed method, the Classical PINN is first introduced.
For solving the Eq.~\eqref{eq:op} with the PINN framework,  a deep neural network is defined to approximate the solution $u$ and enforce the underlying physics of the system. The network takes the form $u_{\theta}(t, x)$, where $\theta$ represents the parameters of the neural network. The loss function for PINN comprises multiple components, capturing the differential equation, initial condition, and boundary condition constraints. Specifically, the loss is given by: 
\begin{equation} 
\begin{aligned} 
\mathcal{L}_{PDE} &= \|\frac{\partial u_{\theta}}{\partial t} - \mathscr{A}(u_{\theta})\|^2_{L^2(0,T;L^2(D))}, \\
\mathcal{L}_{IC} &= \|u_{\theta}(0, x) - u_0(x)\|^2_{L^2(D)}, \\
\mathcal{L}_{BC} &= \|u_{\theta}(t, x) - g(t, x)\|^2_{L^2(0,T;L^2(\partial D))}, \\
\mathcal{L}(\theta) &= 
w_{PDE}\mathcal{L}_{PDE} + w_{IC}\mathcal{L}_{IC} + w_{BC}\mathcal{L}_{BC}, 
\end{aligned} 
\end{equation} 
and three components of the loss function are typically approximated using a suitable numerical quadrature with quadrature points in space-time constituting data sets $\mathcal{S}_{PDE}=\{ (t_{PDE}^i, x_{PDE}^i) \}_{i=1}^{N_{PDE}}$, $\mathcal{S}_{IC}=\{ (t_{IC}^i, x_{IC}^i) \}_{i=1}^{N_{IC}}$, $\mathcal{S}_{BC}=\{ (t_{BC}^i, x_{BC}^i) \}_{i=1}^{N_{BC}}$, $w_{PDE}$, $w_{IC}$ and $w_{BC}$ are weighted to balance the PDE, the initial condition and the boundary condition.
The loss function penalizes deviations from the PDE, initial condition, and boundary condition at their respective data points. In the PDE loss, the exact derivatives are computed using automatic differentiation. The training of the PINN involves minimizing this loss function through optimization algorithms such as stochastic gradient descent or Adam. PINN demonstrate the ability to learn the underlying dynamics of the system, making them a versatile and effective tool for solving complex PDEs. 

\subsection{Spectral PINN}
% space description
To capture the dominant behaviour of the solution for separating the large-scale and small-scale dynamics in the finite-dimensional space, the choice of projection operators $\mathscr{P}_M$ and $\mathscr{Q}_M$ is crucial. Assuming $\{\phi_k\}_{k=1}^N$ as the orthogonal basis for $\mathscr{X}^N$, the following modal decomposition of $u$ can be introduced:
\begin{equation}
u(t,x)=\sum_{k=1}^N U_k(t) \phi_k(x),
\end{equation}
where $U_k(t)$ is determined by the inner product $\langle u,\phi_k\rangle_{\mathscr{X}^N}$.   
Theoretically,  $\mathscr{X}^N$ can be constructed by proper orthogonal decomposition eigenmodes, Fourier modes, orthogonal function spaces, operator normal modes, or Koopman modes, etc.  It's important to note that when representing $u(t,x)$, modal expansion coefficients are randomly distributed in mode order and encompass a broad range of spatial scales. Therefore, spectral expansions (e.g. Fourier modal expansions, Chebyshev Polynomial expansions) are more suitable for this problem.  In this paper, a finite Fourier modal space is utilized to construct the spectral PINN, i.e., $\mathscr{X}^N=\{\exp(\imath k x)\}_{k=1}^N$.  Accordingly, the subspace of large-scale and small-scale can be defined as $\mathscr{X}^M=\{\exp(\imath k x)\}_{k=1}^M$ and ${\mathscr{X}^M}^\perp=\{\exp(\imath k x)\}_{k=M+1}^N$, respectively. And the solutions for Eq.\eqref{eq:ls} and Eq. \eqref{eq:ss} can be represented respectively as:
\begin{equation}
\begin{aligned}
p_M(t,x)&=\sum_{k=1}^M U_k(t) e^{-\imath kx},\\
q_M(t,x)&= \sum_{k=M+1}^N U_k(t) e^{-\imath kx}.
\end{aligned}
\end{equation}
The FFT and inverse FFT are potent tools that could be employed to convert between the spatial domain and frequency domain. So the discretisation of bounded domain $D$ should be considered and then we can define $\mathbf{u}(t)=(\{u(t,x_j)\}_{j=1}^N)$ and $\mathbf{U}(t)=(\{U_k(t)\}_{k=1}^N)$ for convenience.
Then the Eq. \eqref{eq:op}  can be considered in the discrete frequency domain as: 
\begin{equation}
\frac{d \mathbf{U}(t)}{d t}= \mathscr{F} \left(\mathscr{A}\left(\mathscr{F}^{-1}(\mathbf{U}(t))\right)\right).
\end{equation}
The following discussion and the proposed spectral PINN in this paper focus on this frequency equation rather than the original equation in the spatial domain.  The most concerned equation that is used in the loss function of spectral PINN, i.e., the equation for small-scale behaviour,  could be presented in the discrete frequency domain as:
\begin{equation}\label{eq:fss}
\frac{d \breve{\mathbf{U}}^\perp(t)}{d t}= \mathscr{F} \left(\mathscr{Q}_M \mathscr{A}\left(\mathscr{F}^{-1}(\mathbf{U}(t)\right)\right),
\end{equation}
with $\breve{\mathbf{U}}^\perp(t) = (\mathbf{0},\{U_k(t)\}_{k=M+1}^N)$ which could be used to achieve the small-scale solution. Here a zero-vector $\mathbf{0}$ is employed to extend the length to $N$.
For convenience, let  $\breve{\mathbf{U}}(t) = (\{U_k(t)\}_{k=1}^M,\mathbf{0})$  denote the frequency solution set corresponding to the large-scale solution, and it's trivial that $\mathbf{U}(t) = \breve{\mathbf{U}}(t) + \breve{\mathbf{U}}^\perp(t)$.

Given the intricacy of small-scale behaviour, a neural network with a single input, the time $t$, is employed as a tool to solve the Eq. \eqref{eq:fss}. The neural network propagator $\mathcal{G}^\varrho_{\breve{\mathbf{U}}(t)}(t;\theta)$  subjects to the nonlinear equation and the data of specified initial conditions, and the following holds with a proper precision:
\begin{equation}
\breve{\mathbf{U}}^\perp(t) \approx \mathcal{G}^\varrho_{\breve{\mathbf{U}}(t)}(t;\theta).
\end{equation}
The abstract description of $\mathcal{G}^\varrho_{\breve{\mathbf{U}}(t)}(t;\theta)$ is considered as:
\begin{equation}
\mathcal{G}^\varrho_{\breve{\mathbf{U}}(t)}(t;\theta): \mathbb{R}^+ \times \mathbb{R}^{|\theta|} \to {\mathscr{X}^M}^\perp,
\end{equation}
where $\theta$ indicates the set of all parameters, including weights and biases, and $|\theta|$ is the cardinality of $\theta$. $\varrho$ denotes activation functions, and both $\varrho$ and the large-scale physics $\breve{\mathbf{U}}(t)$ should be specified in advance. 
For the sake of brevity, the output of neural network $\mathcal{G}^\varrho_{\breve{\mathbf{U}}(t)}(t;\theta)$ is denoted by $\breve{\mathbf{U}}^\perp_\theta(t)$.

With the output of neural networks, $\breve{\mathbf{U}}^\perp_\theta(t)$,  the loss function used in spectral PINN could be defined as follows:
\begin{equation} 
\begin{aligned} 
\mathcal{L}_{PDE} &= \|\frac{d \breve{\mathbf{U}}^\perp_\theta(t)}{d t} - 
\\
&\mathscr{F}\left(\mathscr{A}\left(\mathscr{F}^{-1}(\breve{\mathbf{U}}^\perp_\theta(t)+\breve{\mathbf{U}}(t))\right)\right)\|^2_{L^2(0,T;l^2)} ,\\
\mathcal{L}_{IC} &= \|\breve{\mathbf{U}}^\perp_\theta(0) - \breve{\mathbf{U}}^\perp_0\|^2_{l^2}  ,\\
\mathcal{L}(\theta) &= w_{PDE}
\mathcal{L}_{PDE} + w_{IC} \mathcal{L}_{IC}, 
\end{aligned} 
\end{equation} 
where $\breve{\mathbf{U}}(t)$ is given by classic numerical methods or neural networks with low cost because it only captures the large-scale behaviour. 

\begin{remark}In the formulation of loss functions for the proposed spectral PINN, only the residuals from the physics-guided nonlinear equations are involved. It is possible to include experimental data in the loss function and even omit the residual part. This means that it can be applied to more complicated problems without a simple equation as guidance, i.e., it can be treated as a purely data-driven process. 
\end{remark}

In computing loss functions, the temporal derivative is typically computed using automatic differentiation for accuracy. However, for high-dimensional problems like three-dimensional Navier-Stokes problems or domains with extensive data or numerous modes, the cost of automatic differentiation can be significant or may even be impractical. In such cases, the finite difference method serves as an alternative for solving the temporal derivative. But for the spatial derivative within the nonlinear operator $\mathscr{A}$,   it can be easily represented in the Fourier space with properties of the Fourier transform:
$$\frac{\partial^n u}{\partial x^n} = \mathscr{F}^{-1}((\imath k)^n \mathscr{F}(u)).$$
So the formulation of the loss functions for spectral PINN avoids the direct computation of spatial derivatives, especially for high-order derivatives, which can significantly reduce computational complexity while ensuring high accuracy. 

\begin{remark}
Classical PINN is designed to learn the global solution point by point in the spatial domain for every time step. This task can involve complicated functions with intricate patterns and interactions in both time and space, especially for systems with chaotic behaviour or turbulence. In contrast to the classical PINN, spectral PINN benefits from the transformation to frequency space, a low-dimensional representation focusing on essential, local frequency information corresponding to the whole space domain, simplifying the learning task. Fourier coefficients, representing the distribution of energy across different frequency components, effectively simplify the spatial complexity and could be more amenable to learning.  According to the energy cascade theory, little whorls are slaved by big whorls and gain energy, a phenomenon associated with frequency rather than specific spatial locations. In addressing such small-scale problems, the spectral PINN is deemed more suitable than classical PINN. 
\end{remark}

\section{Supplementary information}
\textbf{Supplementary Video 1}\\
Supplementary video for the case of 2D NS equations: A comprehensive depiction of the full-scale, extracted large-scale and small-scale fields derived from the numerical simulation at each time step, along with the predicted small-scale dynamics obtained with the proposed method.
\\
\textbf{Supplementary Video 2}\\
Supplementary video for the case of 3D NS equations: A comprehensive depiction of the small-scale fields derived from the numerical simulation at each time step, along with the predicted small-scale dynamics obtained with the proposed method.

\section*{Acknowledgments}
This work is supported by the Natural Science Foundation of China (Nos. 92270109, U23A2069).

\bibliography{ref}

\end{document}